\documentstyle[aps,prl,multicol,epsfig]{revtex}

\begin{document}

\title{Glassy dynamics near zero temperature}

\author{Federico Ricci-Tersenghi\protect\cite{federico} and
	Riccardo Zecchina\protect\cite{riccardo}}

\address{
The Abdus Salam International Center for Theoretical Physics,
Condensed Matter Group\\
Strada Costiera 11, P.O. Box 586, I-34100 Trieste, Italy
}

\maketitle

\begin{abstract}
We numerically study finite-dimensional spin glasses at low and zero
temperature, finding evidences for (i) strong time/space
heterogeneities, (ii) spontaneous time scale separation and (iii)
power law distributions of flipping times. Using zero temperature
dynamics we study blocking, clustering and persistence phenomena.
\end{abstract}


\begin{multicols}{2}
\narrowtext

The study of non-equilibrium dynamics in quenched spin systems
provides a rigorous and numerical test ground for modeling glassy
behavior.  Among the open issues which arise in this context there is
the connection between spatial and temporal heterogeneities and the
global off-equilibrium dynamical features characteristic of glassy
structures such as aging~\cite{YOUNG_BOOK}.  In the last decade, much
progress has been made in the study of domain evolution and aging in
both ordered and disordered systems~\cite{BRAY,AGING,RIEGER}.  In
particular, the mean-field picture of the glassy
transition~\cite{MPV,REVIEW} plays a central role in interpreting
results of numerical simulations on spin systems, but it can not
account for heterogeneities.  On the other hand a set of interesting
numerical and analytical results on the off-equilibrium dynamics in
non-frustrated systems have been obtained~\cite{GODRECHE,DERRIDA,NS},
which can be now extended to frustrated ones.

The scope of this study is to provide a further link between the
different aspects of the glassy transition unfolded by the various
approaches.  More specifically, we shall focus on models with discrete
($\pm J$) couplings in 2 and 3 dimensions.  Of particular interest is
the onset of time scale separation in the low temperature (low $T$)
regime and its connections with the blocking and persistence phenomena
in the zero-temperature ($T\!=\!0$) limit.

The choice of discrete couplings stems from the fact that they
naturally arise as interactions incorporating geometrical frustration
and constraints. Moreover, models with discrete couplings are of
central relevance in other areas such as combinatorial optimization
and are known to provide very rich dynamical features even at
$T\!=\!0$~\cite{BARRAT}.

In what follows, we report results from extensive numerical
simulations concerning the following issues.

{\bf I}) The onset and the dependence on waiting time of the
spontaneous time scale separation and space clustering at sufficiently
low $T$ in 2D and 3D models.

{\bf II}) The related emergence at $T\!=\!0$ of the ``blocking''
phenomenon together with the clustering of a finite fraction of spins
that flip infinitely often (``fast'' spins), along with its
interpretation in terms of the functional mean-field order parameter.

{\bf III}) The $T\!=\!0$ distributions of flipping times and fast
spins clusters sizes.

{\bf IV}) The persistence phenomenon.

Our studies focus on the $\pm J$ Edwards-Anderson (EA) model in 2 and
3 spatial dimensions (square and cubic lattices). The Hamiltonian of
these models reads
\begin{equation}
{\cal H} = \sum_{(i,j)\in E} s_i J_{ij} s_j \qquad ,
\end{equation}
where $E$ is the set of lattice edges, the spins are of Ising type and
the couplings take the values $\pm 1$ with equal probability.

The study of 3D EA model in the very low temperature region requires
huge thermalization times and therefore it has been severely limited
by available computer resources.  Only very recently these studies
have been pushed to very low temperatures, thanks to the use of
parallel computers~\cite{REVIEW}.  The data presented in this letter
referring to the 3D case has been obtained with the help of the APE100
parallel computer~\cite{APE}.  The 2D case is indeed much simpler in
virtue of the existence of polynomial algorithms for ground-states
calculations.

It is known that in both 2D and 3D EA models~\cite{CONIGLIO} lowering
temperature produces a surprising increase in the number of high
frequency flipping spins. Here we push the study of the flipping times
distribution to the very low temperature region in the hard case of
the 3D EA model.  Moreover we investigate the dependence of this
distribution on the waiting time, which is a relevant feature of the
aging regime in glassy phases.

According to the typical scheme used in off-equilibrium dynamical
studies~\cite{REVIEW}, we simulate large systems (of at least $32^3$
spins) and we start the experiment with an instantaneous quench from
infinite temperature to one in the glassy phase ($T < T_c \simeq
1.1$).  Next we let the system evolve for a waiting time $t_w$ (times
are expressed in terms of Monte Carlo sweeps, MCS).  Finally we
calculate the flipping rates probability distribution function (pdf)
measuring the number of flips done by every spin within time windows
extending from $t_w+t$ to $t_w+2t$, where $t=2^k$ (with $k \le 26$) is
also the time window size.  We simply define the mean flipping time
$\tau$ as the time window size divided by the number of flips and we
construct the pdf of its logarithm, $P_{t,t_w}(\ln\tau)$, by taking
the histogram over all spins (with the label $t$ identifying the time
window). This distribution is expected to be self-averaging (like
correlation functions) and so we prefer to simulate few very large
samples.  The choice of working with the $\ln(\tau)$ pdf instead of
that for $\tau$ is dictated by the broadness of the latter.  Note that
a pure exponential tail in the former [$P(\ln\tau) \propto e^{-\lambda
\ln\tau}$] corresponds to a power law tail in the latter [$P(\tau)
\propto \tau^{-\lambda-1}$].

\begin{figure}
\epsfxsize=0.95\columnwidth
\epsffile{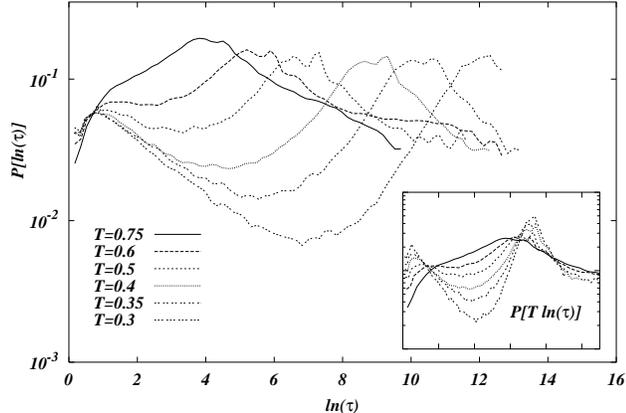}
\caption{Flipping times pdf show a spontaneous time scales separation
as temperature is decreased. In the inset we present the pdf of the
scaling variable $T \ln(\tau)$}
\label{fig1}
\end{figure}

In the $t_w=0$ case we find, as expected, that $P_{t,0}(\ln\tau)$ does
not depend on $t$ and so we consider only the pdf measured during the
last time window, which has the largest support (more than $10^6$ MCS)
and the highest statistics.  In Fig.~\ref{fig1} we show such a
distribution for different temperatures.  While at very high
temperatures the pdf is strongly peaked around the mean, it acquires
very large tails approaching the glass transition
temperature~\cite{CONIGLIO} and finally develops a clear bimodal shape
at very low temperatures, as can be seen in Fig.~\ref{fig1} (note the
logarithmic scale on the $y$ axis).  The emergence of a spontaneous
time scales separation, allows us to naturally divide the spins into
two very general classes: the ``fast'' spins belonging to the left
peak and the ``slow'' ones to the right peak.  While the shape of the
left part of the distribution does not change with
temperature, the position of the ``slow'' peak is clearly moving
towards higher values when the temperature is decreased.  As shown in
the inset of Fig.~\ref{fig1}, this process verify the simple scaling
$T \ln(\tau)$, commonly found in every activated process in spin
glasses.  It follows that slow spins are the ones that have to
overcome a barrier in order to flip, while the fast ones will
eventually have a zero local field at some time (which could even
happen very rarely).  In the zero temperature limit we expect that the
slow peak moves towards unreachable time-scales, whereas fast spins
are the only responsible for the dynamics. The adjective ``fast''
could be somehow misleading, in turn their actual flipping times
follow a very broad distribution.

\begin{figure}
\epsfxsize=0.95\columnwidth
\epsffile{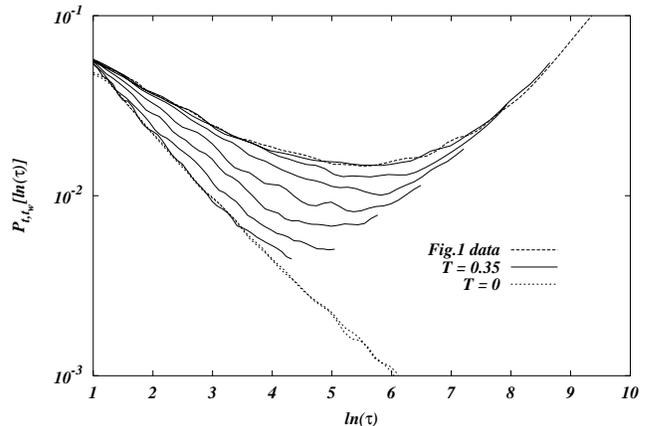}
\caption{$T\!=\!0.35$ data has been measured on a $32^3$ system (10
samples) after waiting $2^{24}$ MCS. Different curves correspond to
different time window sizes (from bottom)
$t=2^{12},2^{14},2^{16},2^{18},2^{20},2^{22},2^{24}$. The last curve
coincides with data presented in Fig.1. $T\!=\!0$ data have been
measured on $100^3$ systems, after a waiting time of
$t_w=2^{16},2^{18}$.}
\label{fig2}
\end{figure}

It is well known that in the aging regime two-times quantities depend
on both times and not only on their difference.  In the case of our
$P_{t,t_w}(\ln\tau)$ this dependence on the waiting time is shown in
Fig.~\ref{fig2}, where we present data for a low temperature
($T\!=\!0.35$), a very large waiting time ($t_w=2^{24}$) and many
values of $t$.  During the aging process two different regimes can be
identified~\cite{AGING,FV}: the {\em quasi-equilibrium} regime ($t \ll
t_w$) where the system relaxes inside a quasi-state and the {\em
aging} regime ($t \ge t_w$) where macroscopic rearrangements take
place.  In these two regimes the shape of the $P_{t,t_w}(\ln\tau)$ is
different (this has been verified in many simulations with different
$t_w$ values, even if here we report the results concerning only one
waiting time).  In Fig.~\ref{fig2} we display the pdf for different
choices of the ratio $t/t_w$: the lowest curves correspond to the
quasi-equilibrium regime, while the upper ones have values of $t$ of
the order of $t_w$ and therefore include the effects coming from the
aging regime.

It is clear that the time-scales separation is sharper in the
quasi-equilibrium regime, while approaching the aging regime the gap
between fast and slow peaks is partially filled.  Observing that the
slow spins create the environment where the fast spins move in (the so
called cage effect), we may assume that the meta-stable state is
identified by the spatial structure of slow spins which act as a rigid
backbone on time scales typical of the quasi-equilibrium regime.
Therefore we may also observe that the basic excitations that bring
the system out of quasi-equilibrium are given by those spins which
fill the gap between the peaks.  The study of the spatial structure of
these spins would help towards the understanding of basic excitation
in 3D spin glasses.

The same set of simulations with Gaussian couplings gives no evidence
of a so simple time scale separation.  In this case the distribution
of microscopic time scales is different and much longer simulations
would be needed in order to unveil the separation.  Moreover at
$T\!=\!0$ in the Gaussian case the system rapidly get stuck in a local
minimum and its configuration is fully frozen, consistently with the
zero-temperature classification given in~\cite{NS}.  Only the $\pm J$
model has a rich aging behavior at zero temperature.

In the $T\!=\!0$ limit activated processes disappear and the slow peak
goes to infinity.  Then the gap between the peaks can no longer be
filled when the system leaves the quasi-equilibrium regime.  In other
words, the $P_{t,t_w}(\ln\tau)$ becomes $t_w$-independent (we have
checked numerically this fact) and it describes only the fast spin
component.  Remarkably enough, the $T\!=\!0$ pdf is already present in
the finite temperature data, in their quasi-equilibrium regime (see
Fig.~\ref{fig2}) and this would suggest that $T\!=\!0$ dynamical
features could be present at low temperatures too, at least in the
quasi-equilibrium regime.  Hereafter we will consider only the
following updating rule. A site is randomly chosen and it is oriented
in the direction of the neighbors majority or at random if the local
field on it is zero.

We find that the $T\!=\!0$ flipping time distributions can be well
fitted by a power law over many decades, with an exponent
$\lambda+1=1.76(2)$ in 3D and $\lambda+1=2.08(3)$ in 2D.  This means
that in 2D after a large but {\em finite} time (given by the average
flipping time) an extensive fraction of the system can be flipped with
zero-energy costs.  On the contrary in 3D an average flipping time can
not be defined and this divergence is maybe related to the existence
of a finite temperature phase transition.

{\it Persistence}~\cite{GODRECHE} is one the most studied properties
of zero temperature dynamics~\cite{DERRIDA,NS}, nevertheless it has
been never measured in spin glasses. Here we fill this gap.

The persistence $U(t,t_w)$ is defined as the number of unflipped spins
in the time interval $[t_w,t_w+t]$.  In pure ferromagnetic models (at
least for $D=2,3$) the persistence $U(t,t_w)$ decays to zero, in the
large times limit, as $U(t,t_w) \propto (t/t_w)^{-\theta(D)}$, the
exponent being $t_w$-independent.  On the contrary, in disordered
models the persistence is expected to remain finite~\cite{NS} in the
$t\to\infty$ limit and its asymptotic value $U_\infty(t_w)$ may depend
on the waiting time.  The physical mechanism underlying such effect is
the slow freezing of a rigid component.  It follows that the proper
way of estimating the persistence in disordered systems is by taking a
sufficiently large value of $t_w$, differently from what has been done
in previous studies where the choice $t_w=0$ was used.

\begin{figure}
\epsfxsize=0.95\columnwidth
\epsffile{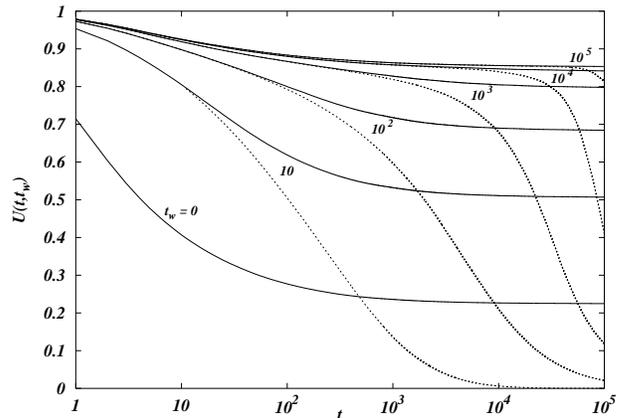}
\caption{Continuous lines show the persistence decay after different
waiting times in a $100^3$ spin glass. Dashed lines present the
persistence in a microcanonical experiment, where the energy is kept
constant after time $t_w$.}
\label{fig3}
\end{figure}

In Fig.~\ref{fig3} we show the results for 3D $\pm J$ spin glasses
(very similar results have been found in the 2D case).  While the
system relaxes towards the stationary state the number of frozen spins
grows and $U_\infty(t_w)$ monotonically increases with $t_w$. In the
large times limit we can extract the $\theta$ exponent from the decay
$U(t,t_w) - U_\infty(t_w) \propto t^{-\theta}$.  This fitting
procedure is not an easy one, because of the slow persistence decay,
and so our results are affected by large errors.  Our best estimations
are $\theta=0.46(3)$ and $U_\infty(\infty)=0.862(2)$ in 3D and
$\theta=0.64(3)$ and $U_\infty(\infty)=0.778(2)$ in 2D.  The $\theta$
value in 3D is confirmed by the local magnetizations pdf which can be
well fitted by the law $(1-m^2)^{\theta-1}$ at low and zero
temperatures.

These numbers imply that in the large times limit a $\pm J$ spin glass
is not completely frozen.  While a large fraction $U_\infty(\infty)$
of the spins get blocked forever, a {\it finite} fraction
$1-U_\infty(\infty)$ of the volume is composed by spins that flip
infinitely often~\cite{GNS}, with their flipping times following power
law distributions. The fast spins produce the persistence phenomenon.

In Fig.~\ref{fig3} we report with dashed lines the results of a
microcanonical experiment. As before, we let the system relax for
$t_w$ MCS and then we measure the persistence. The difference is that
now the dynamics after time $t_w$ is performed keeping the energy
constant to the value it had at time $t_w$. For $t \le t_w$ the data
perfectly coincide with those for the usual persistence.  Then the
system leaves the configurational space region (state~\cite{REVIEW})
where it was confined by the dynamics and it goes far away (we check
this also measuring correlations functions, that behave qualitatively
like persistence).  Thus we can conclude that the freezing phenomenon
is completely dynamical~\cite{note2}.  In other words the system seems
to be stuck only because it always finds a path towards a lower energy
configuration belonging to the same state before finding a path
towards a different state.

For both dimensionalities we find that fast spins are organized in
clusters.  At any time and in each cluster there must be at least one
``free'' spin that can be flipped without any energy cost ($\Delta E
\le 0$).  The number of free spins equals the number of flat
directions in the configuration space and decreases lowering the
energy.  Free spins usually appear in pairs, which wander inside
their cluster~\cite{note1}.

\begin{figure}
\epsfxsize=0.95\columnwidth
\epsffile{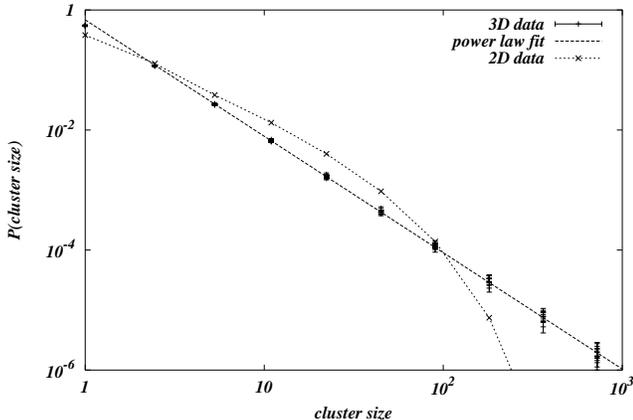}
\caption{Fast spins clusters sizes pdf. 3D data have been collected in
system of sizes $30^3$, $45^3$ and $100^3$, and with two different
updating rules (sequential and random). The power law fit gives
$-1.94(3)$ as the best exponent. 2D data, measured in a $1000^2$
system, clearly show a cut-off.}
\label{fig4}
\end{figure}

Fast spins clusters sizes follow the pdf shown in Fig.~\ref{fig4}.
Measurements have been taken after a waiting time of $t_w=10^4$. Note
that a finite $t_w$ can overestimate a little the clusters size,
however for both dimensions we have that $U_\infty(10^4)$ is
very near to $U_\infty(\infty)$ and we did not find any dependence on
$t_w$.  Up to our accuracy, 3D data can be well fitted by a power law
with an exponent $-1.94(3)$.  On the contrary 2D data clearly show a
cut-off.  We can not exclude that also 3D data would show a cut-off on
larger scales.

For any finite cluster we can construct the set of allowed
configurations and then we can evaluate the time a random walker needs
to visit all of them. This time is, by construction, greater than the
larger flipping time of any spin in the cluster.  In this way we can
link clusters sizes and flipping times, and one would expect to find a
cut-off $\tau_{max}$ in the 2D distribution of flipping times as a
consequence of the one present at $c_{max}$ in Fig.~\ref{fig4}.
However times may be exponential in the cluster size (like e.g.\ the
recurrence time) and $\tau_{max}$ may be considered infinite for all
practical purposes.  Moreover, if one assume from percolation
arguments that $c_{max}$ grows logarithmically with the system size
$N$, then $\tau_{max}$ would grow as a power of $N$, like usually in
any finite size systems.

In conclusion we have performed a numerical study of the low
temperature dynamics of a finite-dimensional frustrated model with
discrete couplings, namely the EA model in 2D and 3D, which can be
viewed as a prototype model for glassy systems.  The onset of a
spontaneous time/space scale separation allows for a classification of
spins in distinct classes characterized by their own dynamics as well
as the identification of low-energy excitation responsible for
structural rearrangements during the aging.

In the $T\!=\!0$ limit, while a part of the system completely freeze,
the fast spins dynamics can be quantitatively described in terms of
power law exponents which distinguish the 2D from the 3D case.  The
dynamics inside a cluster of fast spins, although diffusive-like, has
strong constraints and can be interpreted along the lines
of~\cite{PSAA}.

Given a proper definition of persistence in aging systems, we have
been able to estimate the persistence exponent in the EA model, which
turns out to be of kind ${\cal M}$~\cite{NS}, i.e.\ partially frozen
and partially infinitely flipping.  Finally, with a microcanonical
experiment, we have shown that blocking phenomena are completely
dynamical, even in disordered systems.

In spite of the central role played by space and time heterogeneities
in glassy dynamics, such aspects have been rather poorly studied from
the analytical point of view.  Scope of this paper has been to provide
some reference numerical result which hopefully could be captured by
theoretical arguments.  A first step in this direction can be done
within the framework of finitely connected long range models (like
e.g.\ the Viana-Bray one~\cite{VB}).  In these models the underlying
graph generates non trivial heterogeneities, driven by the
connectivity pattern.

\end{multicols}

\vspace{-.5cm}
\begin{thebibliography}{99}
\vspace{-1.5cm}

\bibitem[\dag]{federico} E-mail: {\tt riccife@ictp.trieste.it}

\bibitem[\ddag]{riccardo} E-mail: {\tt zecchina@ictp.trieste.it}

\bibitem{YOUNG_BOOK} {\it Spin Glasses and Random Fields}, edited by
A.P.~Young. World Scientific (Singapore 1997).

\bibitem{BRAY} A. J. Bray, Adv. Phys. {\bf 43}, 357 (1994).

\bibitem{AGING} J.P. Bouchaud, L.F. Cugliandolo, J. Kurchan and
M. M\'ezard in~\protect\cite{YOUNG_BOOK}, pag.~161.

\bibitem{RIEGER} H. Rieger, Ann. Rev. Comp. Phys. {\bf 2}, 295 (1995).

\bibitem{MPV} M. M\'ezard, G. Parisi, M. A. Virasoro, {\em Spin glass
theory and beyond} (World Scientific, 1987).

\bibitem{REVIEW} E. Marinari, G. Parisi, F. Ricci-Tersenghi,
J.J.~Ruiz-Lorenzo and F.~Zuliani, J. Stat. Phys. {\bf 98}, 973 (2000).

\bibitem{GODRECHE} A.J. Bray, B. Derrida and C. Godr\`eche,
Europhys. Lett. {\bf 27}, 175 (1994); J. Phys. A {\bf 27}, L357
(1994).

\bibitem{DERRIDA} B. Derrida, V. Hakim and V. Pasquier,
Phys. Rev. Lett.  {\bf 75}, 751 (1995); J. Stat. Phys. {\bf 85}, 763
(1996).

\bibitem{NS} C.M. Newman and D.L. Stein, Phys. Rev. Lett. {\bf 82},
3944 (1999); Physica A {\bf 279}, 159 (2000).

\bibitem{BARRAT} A. Barrat and R. Zecchina, Phys. Rev. E {\bf 59},
1299 (1999).

\bibitem{APE} C. Battista {\it et al.}, Int. J. High Speed Comp. {\bf
5}, 637 (1993).

\bibitem{CONIGLIO} P.H. Poole, S.C. Glotzer, A. Coniglio and N. Jan,
Phys. Rev. Lett. {\bf 78}, 3394 (1997).

\bibitem{FV} S. Franz and M.A. Virasoro, J. Phys. A {\bf 33}, 891
(2000).

\bibitem{GNS} The same happens in a 2D spin glass too. For a proof see
A. Gandolfi, C.M.  Newman and D.L. Stein, to appear in Communications
in Mathematical Physics (2000).

\bibitem{note2} This conclusion is trivial for non-disordered models
where domain walls can freely wander, but it was not so for disordered
models where domain walls are pinned.

\bibitem{note1} The movement of a pair is induced by the flip of one
of its two free spins.  The precise characterization of this diffusive
dynamics (whether normal, sub- or super-diffusive) requires further
studies.

\bibitem{PSAA} R.G. Palmer, D.L. Stein, E. Abrahams and P.W. Anderson,
Phys. Rev. Lett. {\bf 53}, 958 (1984).

\bibitem{VB} L. Viana and A.J. Bray, J. Phys. C {\bf 18}, 3037 (1985).
M. M\'ezard and G. Parisi, Europhys. Lett. {\bf 3}, 1067 (1987).
I. Kanter and H. Sompolinsky, Phys. Rev. Lett. {\bf 58}, 164 (1987).

\end{thebibliography}
\end{document}